\documentclass[12pt, reqno]{amsart}
\usepackage{amssymb}
\usepackage{mathrsfs}
\usepackage{mathpazo}
\usepackage[colorlinks=true, breaklinks=true]{hyperref}
\usepackage{bm}
\usepackage[latin1]{inputenc}
\usepackage{graphicx,epsfig,graphics}
\usepackage{subfigure}
\newtheorem{theorem}{Theorem}

\newtheorem{corollary}[theorem]{Corollary}

\newtheorem{lemma}[theorem]{Lemma}

\newtheorem{proposition}[theorem]{Proposition}
\newtheorem{remark}[theorem]{Remark}

\newcommand{\bea}{\begin{eqnarray}}
\newcommand{\eq}{\end{eqnarray}}
\newcommand{\eea}{\end{eqnarray}}
\newcommand{\bqn}{\begin{eqnarray*}}
\newcommand{\beaa}{\begin{eqnarray*}}
\newcommand{\eqn}{\end{eqnarray*}}
\newcommand{\eeaa}{\end{eqnarray*}}
\newcommand{\bpr}{\begin{proposition}}
\newcommand{\epr}{\end{proposition}}

\newcommand{\cal}{\mathcal}

\title{Left-wing asymptotics of the implied volatility in the presence of atoms}
\author{Archil Gulisashvili}
\address{Department of Mathematics, Ohio University}
\email{gulisash@ohio.edu}
\date{}

\begin{document}

\begin{abstract}
We consider the asymptotic behavior of the implied volatility in stochastic asset price models with atoms. In such models, the asset price distribution has a singular component at zero. Examples of models with atoms include the constant elasticity of variance model, jump-to-default models, and stochastic models described by processes stopped at the first hitting time of zero. For models with atoms, the behavior of the implied volatility at large strikes is similar to that in models without atoms. On the other hand, the behavior of the implied volatility at small strikes is influenced significantly by the atom at zero. S. De Marco, C. Hillairet, and A. Jacquier found an asymptotic formula for the implied volatility at small strikes with two terms and also provided an incomplete description of the third term. In the present paper, we obtain a new asymptotic formula for the left wing of the implied volatility, which is qualitatively different from the De Marco-Hillairet-Jacquier formula. The new formula contains three explicit terms and an error estimate. We show how to derive the De Marco-Hillairet-Jacquier formula from our formula, and compare the performance of the two formulas in the case of the CEV model. The resulting graphs show that the new formula provides a notably better approximation to the smile in the CEV model than the De Marco-Hillairet-Jacquier formula.
\end{abstract}
\maketitle

\section{Introduction}\label{S:intro}
During the last decade, several significant model-free formulas, describing the asymptotic behavior of the implied volatility at extreme strikes, were found. We only mention here R. Lee's moment formulas (see \cite{L}), tail-wing formulas due to S. Benaim and P. Friz (see \cite{BFa,BFb}, see also \cite{BFL}), asymptotic formulas with error estimates established by the author (see \cite{Gb,Gc}), and 
\let\thefootnote\relax\footnote{The author thanks Antoine Jacquier and Stefano De Marco for reading the paper and making valuable comments. The author is also indebted to Antoine Jacquier for providing the graphs included in the last section of the paper.}higher order formulas found by K. Gao and R. Lee (see \cite{GL}). We refer the interested reader to the book \cite{Ga} by the author for more information.

The present work was inspired by the paper \cite{DMHJ} of S. De Marco, C. Hillairet, and A. Jacquier. The authors of \cite{DMHJ} obtained interesting results concerning the asymptotic behavior of the implied volatility at small strikes in the case where the asset price distribution has an atom at zero (see Theorem 3.7 in \cite{DMHJ}). Special examples of such models are the constant elasticity of variance model, jump-to-default models, and stochastic models described by processes stopped at the first hitting time of zero (more information can be found in \cite{DMHJ}). It is not hard to see that the right-wing behavior of the implied volatility in models with and without atoms is similar. Therefore, general model-free asymptotic formulas for the implied volatility at large strikes, discussed in Chapter 9 of \cite{Ga}, can be used 
in stochastic asset price models with atoms. However, the left-wing behavior of the implied volatility in models with and without atoms is qualitatively different. This fact was noticed and explored in \cite{DMHJ}. It was also shown in \cite{DMHJ} that the general formula formulated in Corollary 9.31 in \cite{Ga}, which describes the left-wing behavior of the implied volatility in terms of the put pricing function, holds for asset price models with atoms (see formula (\ref{E:A}) below). For such models, the above-mentioned formula provides only the leading term in the asymptotic expansion of the implied volatility and an error estimate. The authors of \cite{DMHJ} found a sharper asymptotic formula, characterizing the left-wing behavior of the implied volatility in models with atoms (see formula (\ref{E:one}) below). Note that the impact of an atom at zero on the left-wing asymptotics of the implied volatility was not taken into account in
Section 9.9 of \cite{Ga}. This omission led to an incorrect description of the asymptotic behavior of the implied volatility at small strikes in the CEV model (see formula (11.22) in Theorem 11.5 in \cite{Ga}). Only the absolutely continuous part of the distribution of the asset price was taken into account in formula (11.22) mentioned above, while the influence of the atom at zero was ignored. 

In this paper, we establish new asymptotic formulas for the implied volatility at small strikes in models with atoms
(see formulas (\ref{E:imvol2}), (\ref{E:imvol222}), and (\ref{E:imvol22}) below). These formulas contain three explicit terms in the asymptotic expansion of the implied volatility and an error estimate. Note that the asymptotic formula found in \cite{DMHJ} contains two terms and only an incomplete information about the third term is provided. Moreover, there is a qualitative difference between the new formulas and the De Marco-Hillairet-Jacquer formula. In the new formulas, we use the inverse function of a strike-dependent function, while the inverse function of the cumulative standard normal distribution function is employed in \cite{DMHJ}. It is shown numerically in Section \ref{S:num} of the present paper that formula (\ref{E:imvol22}) provides a significantly better approximation to the left wing of the implied volatility in the constant elasticity of variance model than the De Marco-Hillairet-Jacquier formula.

Our next goal is to introduce several known objects, which will be used in the rest of the paper, and then formulate our main results. The asset price will be modeled by a non-negative martingale $X$ defined on a filtered probability space
$(\Omega,{\cal F},\{{\cal F}_t\},\mathbb{P})$. The initial condition for the process $X$ is denoted by $x_0$, and it is assumed that $x_0$ is a positive number. It is also assumed that the interest rate is equal to zero. In the sequel, the symbols $C$ and $P$ stand for the call and put pricing functions, associated with the price process $X$. These functions are defined 
as follows: 
$$
C(T,K)=\mathbb{E}\left[(X_T-K)^{+}\right]\quad\mbox{and}\quad P(T,K)=\mathbb{E}\left[(K-X_T)^{+}\right].
$$
In the previous formulas, $K$ is the strike price, and $T$ is the maturity. The implied volatility $I_C$ is the function, satisfying the following condition:
\begin{equation}
C_{BS}(T,K,I_C(T,K))=C(T,K).
\label{E:iv}
\end{equation}
The expression on the left-hand side of (\ref{E:iv}) is the call pricing function in the Black-Scholes model with the volatility parameter equal to  $I_C(T,K)$. The function $C_{BS}$ is defined by
\begin{equation}
C_{BS}(T,K,\sigma)=x_0{\cal N}(d_1(T,K,\sigma))-K{\cal N}(d_2(T,K,\sigma)),
\label{E:cumul}
\end{equation}
where ${\cal N}$ is the standard normal cumulative distribution function, 
that is, the function
$$
{\cal N}(x)=\frac{1}{\sqrt{2\pi}}\int_{-\infty}^xe^{-\frac{y^2}{2}}dy.
$$
The functions $d_1$ and $d_2$ in (\ref{E:cumul}) are defined by
$$
d_1(T,K,\sigma)=\frac{\log x_0-\log K+\frac{1}{2}\sigma^2T}{\sigma\sqrt{T}}
$$
and
$$
d_2(T,K,\sigma)=\frac{\log x_0-\log K-\frac{1}{2}\sigma^2T}{\sigma\sqrt{T}}.
$$
\begin{remark}\label{R:r} \rm
It will be assumed throughout the paper that $C(T,K)> 0$ for all $T> 0$ and $K\ge x_0$. Moreover,
we assume that $P(T,K)> 0$ for all $T> 0$ and $K< x_0$. The first of the previous restrictions guarantees that $I_C(K)$ exists for all $K\ge x_0$, while under the second restriction, the implied volatility $I_C(K)$ exists for all $K< 1$ (more details can be found in Section 9.1 of \cite{Ga}).
\end{remark}
\begin{remark}\label{R:11} \rm 
The maturity $T> 0$ will be fixed throughout the paper. To simplify notation, we will suppress the symbol $T$ in the function $C$ and in similar functions.
\end{remark}
\begin{remark}\label{R:navigate} \rm
In the proofs of the results obtained in the present paper, we often assume that $x_0=1$. It is easy to understand why this assumption does not restrict the generality. Indeed, let us define a new stochastic process by $\widetilde{X}=x_0^{-1}X$, and denote the corresponding call pricing function and the implied volatility by $\widetilde{C}$ and $\widetilde{I}_{\widetilde{C}}$, respectively. Then, it is not hard to see that $C(K)=x_0\widetilde{C}\left(x_0^{-1}K\right)$. Moreover, the same formula holds for the Black-Sholes call pricing function, and therefore
\begin{equation}
I_C(K)=I_{\widetilde{C}}\left(x_0^{-1}K\right).
\label{E:nownot}
\end{equation}
Since the process $\widetilde{X}$ has $\widetilde{x}_0=1$ as its initial condition,
formula (\ref{E:nownot}) allows us to navigate between asymptotic formulas for the implied volatility under the restriction $x_0=1$ and similar formulas in the general case.
\end{remark}

Let $T> 0$, $x_0=1$, and set
$m_T=\mathbb{P}(X_T=0)$.
If for every $T> 0$, we have $m_T=0$, then the function $G$ defined by
\begin{equation}
G(T,K)=KP\left(T,K^{-1}\right)
\label{E:G}
\end{equation}
is a call pricing function (see \cite{Ga}). The function $G$ plays the role of a link between the left-wing and the right-wing asymptotics of the implied volatility (see \cite{Ga}). 

Now, suppose $0<m_T< 1$ for some $T> 0$. Let us fix such a maturity $T$, and consider $C$, $P$, and $I_C$ as functions of the strike price $K$. Note that for models with atoms, the function $G$, given by (\ref{E:G}), is not a call pricing function. Indeed, the function $G$ does not satisfy the condition $G(K)\rightarrow 0$ as $K\rightarrow\infty$. However, the function $G$ has many features of a call pricing function. For example, it is not hard to see that the Black-Scholes implied volatility $I_G$ exists for all $K\ge 1$ (see Remark \ref{R:r}), and, in addition,
\begin{equation}
I_C\left(K\right)=I_G\left(K^{-1}\right)
\label{E:r1}
\end{equation}
for all $K$ with $0< K\le 1$. We also have
\begin{equation}
G(K)=m_T+\psi(K),
\label{E:r2}
\end{equation}
where $\psi$ is a positive function such that 
\begin{equation}
\psi(K)\rightarrow 0\quad\mbox{as}\quad K\rightarrow\infty. 
\label{E:m}
\end{equation}

The proof of (\ref{E:r2}) and (\ref{E:m}) is simple. Indeed, it follows from 
the definition of the put pricing function that 
$$
G(K)=m_T+\left[\int_0^{\frac{1}{K}}d\widetilde{\mu}_T(x)
-K\int_0^{\frac{1}{K}}xd\widetilde{\mu}_T(x)\right],
$$
where $\widetilde{\mu}_T$ is the distribution of $X_T$ on the open half-line $(0,\infty)$. Hence,
\begin{equation}
\psi(K)=\int_0^{\frac{1}{K}}d\widetilde{\mu}_T(x)
-K\int_0^{\frac{1}{K}}xd\widetilde{\mu}_T(x)\le\int_0^{\frac{1}{K}}d\widetilde{\mu}_T(x).
\label{E:lim}
\end{equation}
Now it is clear that (\ref{E:lim}) implies (\ref{E:r2}). Finally, the proof of the equality in (\ref{E:r1}) 
for asset price models with atoms is the same as the proof of Lemma 9.23 in \cite{Ga}. 
\begin{remark}\label{R:distd} \rm We denote by $p_T$ and $\widetilde{p}_T$ the cumulative distribution functions of the random variable $X_T$ on $[0,\infty)$ and $(0,\infty)$, respectively. These functions are given by
$$
p_T(u)=\mu_T([0,u))\quad\mbox{and}\quad\widetilde{p}_T(u)=\widetilde{\mu}_T([0,u)).
$$
It is clear that
$$
p_T(K)=m_T+\int_0^Kd\widetilde{\mu}_T(x),\quad 0< K<\infty,
$$
$$
\widetilde{p}_T(K)=\int_0^Kd\widetilde{\mu}_T(x),\quad 0< K<\infty.
$$
\end{remark}

We have already mentioned the asymptotic formula in Corollary 9.31 in \cite{Ga}. This formula is as follows:
\begin{align}
I_C(K)&=\frac{\sqrt{2}}{\sqrt{T}}\left[\sqrt{\log\frac{1}{P(K)}}-\sqrt{\log\frac{K}{P(K)}}\right] \nonumber \\
&\quad+O\left(\left(\log\frac{K}{P(K)}\right)^{-\frac{1}{2}}\log\log\frac{K}{P(K)}\right)
\label{E:A}
\end{align}
as $K\rightarrow 0$. Using the mean value theorem and the formula 
$$
P(K)=Km_T+K\psi\left(K^{-1}\right),
$$ 
for small values of $K$ (the previous formula follows from (\ref{E:G}) and (\ref{E:r2})), we obtain
\begin{equation}
I_C(K)=\frac{\sqrt{2}}{\sqrt{T}}
\sqrt{\log\frac{1}{K}}+O(1)
\label{E:zero}
\end{equation}
as $K\rightarrow 0$. This means that in the presence of atoms, the general formula given in (\ref{E:A}) provides only the leading term in the asymptotic expansion of the implied volatility near zero. The expression for the leading term given in (\ref{E:zero}) can also be predicted from Lee's moment formula (see \cite{L}). Indeed, for models with atoms, all the moments of negative order of the distribution of the asset price $X_T$ explode.

We will next formulate the main result of \cite{DMHJ}, adapting it to our notation. Suppose there exists $\varepsilon> 0$ such that
$$
\widetilde{p}_T(u)=O\left(u^{\varepsilon}\right)
$$
as $u\rightarrow 0$. Then
\begin{align}
I_C(K)&=\frac{\sqrt{2}}{\sqrt{T}}\sqrt{\log\frac{1}{K}}+\frac{{\cal N}^{-1}(m_t)}{\sqrt{T}}
+\frac{\sqrt{2}{\cal N}^{-1}(m_t)^2}{4\sqrt{T}\sqrt{\log\frac{1}{K}}} \nonumber \\
&\quad+\Phi(K)
\label{E:one}
\end{align}
as $K\rightarrow 0$. In (\ref{E:one}), the symbol ${\cal N}^{-1}$ stands for the inverse function of the standard normal cumulative distribution function ${\cal N}$. 
In addition, the function $\Phi$ in (\ref{E:one}) satisfies a special estimate of order $O\left(\left(\log\frac{1}{K}\right)^{-\frac{1}{2}}\right)$ as $K\rightarrow 0$ (see Theorem 3.7 in \cite{DMHJ} for more details). Note that the term
\begin{equation}
\frac{\sqrt{2}{\cal N}^{-1}(m_t)^2}{4\sqrt{\log\frac{1}{K}}}
\label{E:inter}
\end{equation}
in (\ref{E:one}) is not really the third term in the asymptotic expansion of $I_C$, because of an interplay 
between the expression in (\ref{E:inter}) and the function $\Phi(K)$ as $K\rightarrow 0$.

In Theorem \ref{T:corrf} and Corollary \ref{C:sims} below, we provide asymptotic formulas for the implied volatility $I_C$ at small strikes with three terms and error estimates of order
$O\left(\left(\log\frac{1}{K}\right)^{-\frac{3}{2}}\right)$ as $K\rightarrow 0$. The main novelty in our approach is that instead of the function ${\cal N}^{-1}$ used in formula (\ref{E:one}), we employ a family of strike-dependent inverse functions $(U_K)^{-1}$, $K> 1$, where  
\begin{equation}
U_K(x)={\cal N}(x)-\frac{1}{2\sqrt{\pi}\sqrt{\log K}}e^{-\frac{x^2}{2}},\quad x\in\mathbb{R}. 
\label{E:fina3}
\end{equation}
It is easy to see that for every $K> 1$, the function $U_K$ is strictly increasing on the interval $[-\sqrt{2}\sqrt{\log K},\infty)$ (differentiate!),
and moreover
$$
\lim_{x\rightarrow-\infty}U_K(x)=0\quad\mbox{and}\quad\lim_{x\rightarrow\infty}U_K(x)=1.
$$
It follows from the reasoning above that the inverse function $(U_K)^{-1}$ exists for all $y$ with 
\begin{equation}
U_K(-\sqrt{2}\sqrt{\log K})\le y< 1,
\label{E:mainr}
\end{equation} 
is strictly increasing on the interval $[U_K(-\sqrt{2}\sqrt{\log K}),1)$, and maps this interval onto the interval $[-\sqrt{2}\sqrt{\log K},\infty)$. 
\begin{remark}\label{R:notethat} \rm
Note that $(U_K)^{-1}(y)$ exists for all $y$ such that
\begin{equation}
{\cal N}(-\sqrt{2}\sqrt{\log K})\le y< 1,
\label{E:accord}
\end{equation}
and therefore,
$(U_K)^{-1}(y)$ exists for all $y$ with $\frac{1}{2}\le y< 1$. On the other hand, if $0< y<\frac{1}{2}$,
then we have to assume that condition (\ref{E:mainr}) holds.
\end{remark}

According to Remark \ref{R:notethat}, $(U_K)^{-1}(m_T)$ is defined for all $K> 1$ provided that $\frac{1}{2}\le m_T< 1$. 
In addition, if $0< m_T<\frac{1}{2}$, then $(U_K)^{-1}(m_T)$ is defined under the following restriction:
\begin{equation}
U_K(-\sqrt{2}\sqrt{\log K})\le m_T.
\label{E:acco1}
\end{equation}
Note that, given $m_T$, there exists $\widetilde{K}> 1$ such that (\ref{E:acco1}) holds for all $K>\widetilde{K}$
and moreover $0< m_T\le G(K)< 1$. Therefore, for all $K>\widetilde{K}$,
we can solve the equations
$U_K(x)=G(K)$ and $U_K(x)=m_T$ by inverting the function $U_K$. 

The next lemma is simple, and we omit the proof.
\begin{lemma}\label{L:analysis} 
Let $K> 1$, and suppose $y$ satisfies (\ref{E:accord}).
Then the inequality $(U_K)^{-1}(y)> 0$ holds if and only if
$$
y>\frac{1}{2}-\frac{1}{2\sqrt{\pi}\sqrt{\log K}}.
$$
\end{lemma}

The next two statements are the main results of the present paper.
\begin{theorem}\label{T:corrf}
Let $x_0> 0$. Then the following asymptotic formula holds for the implied volatility $I_C(K)$ in the asset price models such as above:
\begin{align}
&I_C(K)=\frac{\sqrt{2}}{\sqrt{T}}\left(\log\frac{x_0}{K}\right)^{\frac{1}{2}}
+\frac{1}{\sqrt{T}}\left(U_{\frac{x_0}{K}}\right)^{-1}\left(G\left(\frac{x_0}{K}\right)\right) \nonumber \\
&\quad+\frac{\sqrt{2}}{4\sqrt{T}}\left[\left(U_{\frac{x_0}{K}}\right)^{-1}\left(G\left(\frac{x_0}{K}\right)\right)\right]^2
\left(\log\frac{x_0}{K}\right)^{-\frac{1}{2}} \nonumber \\
&\quad+O\left(\left(\log\frac{x_0}{K}\right)^{-\frac{3}{2}}\right)
\label{E:imvol2}
\end{align}
as $K\rightarrow 0$.
\end{theorem}
\begin{corollary}\label{C:simson}
Let $x_0> 0$, and suppose the random variable $X_T$ is such that
\begin{equation}
\widetilde{p}_T(K)=O\left(\left(\log\frac{1}{K}\right)^{-\frac{3}{2}}\right)
\label{E:ost}
\end{equation}
as $K\rightarrow 0$. Then
\begin{align}
&I_C(K)=\frac{\sqrt{2}}{\sqrt{T}}\left(\log\frac{x_0}{K}\right)^{\frac{1}{2}}
+\frac{1}{\sqrt{T}}\left(U_{\frac{x_0}{K}}\right)^{-1}\left(p_T\left(\frac{K}{x_0}\right)\right) \nonumber \\
&\quad+\frac{\sqrt{2}}{4\sqrt{T}}\left[\left(U_{\frac{x_0}{K}}\right)^{-1}\left(p_T\left(\frac{K}{x_0}\right)\right)\right]^2
\left(\log\frac{x_0}{K}\right)^{-\frac{1}{2}} \nonumber \\
&\quad+O\left(\left(\log\frac{x_0}{K}\right)^{-\frac{3}{2}}\right)
\label{E:imvol222}
\end{align}
as $K\rightarrow 0$.
\end{corollary}
\begin{corollary}\label{C:sims}
Let $x_0> 0$, and suppose the random variable $X_T$ satisfies condition (\ref{E:ost}). Then
\begin{align}
&I_C(K)=\frac{\sqrt{2}}{\sqrt{T}}\left(\log\frac{x_0}{K}\right)^{\frac{1}{2}}
+\frac{1}{\sqrt{T}}\left(U_{\frac{x_0}{K}}\right)^{-1}\left(m_T\right) \nonumber \\
&\quad+\frac{\sqrt{2}}{4\sqrt{T}}\left[\left(U_{\frac{x_0}{K}}\right)^{-1}\left(m_T\right)\right]^2
\left(\log\frac{x_0}{K}\right)^{-\frac{1}{2}} \nonumber \\
&\quad+O\left(\left(\log\frac{x_0}{K}\right)^{-\frac{3}{2}}\right)
\label{E:imvol22}
\end{align}
as $K\rightarrow 0$.
\end{corollary}

It is not hard to see that Corollary \ref{C:sims} follows from Theorem \ref{T:corrf}, Remark \ref{R:distd},  formulas (\ref{E:r2}) and (\ref{E:lim}), and the mean value theorem.
\begin{remark}\label{R:vdrug} \rm
It is interesting to notice that in formulas (\ref{E:imvol2}) and (\ref{E:imvol22}), the term of order $(\log K)^{-1}$
is absent. This happens because of certain cancellations, which occur when we combine formulas (\ref{E:ex1}) 
and (\ref{E:imvol3}) in the proof of Theorem \ref{T:corrf}.
\end{remark}
\begin{remark}\label{R:between} \rm The smile approximations provided in formulas (\ref{E:imvol2}) and (\ref{E:imvol222}) 
take into account the distribution $\mu_T$ of the asset price $X_T$, while the approximations in formula 
(\ref{E:imvol22}) and in the De Marco-Hillairet-Jacquier formula (see (\ref{E:one})) use only the mass $m_T$ 
of the atom at zero.
\end{remark}
\begin{remark}\label{R:vdrugs} \rm Comparing the De Marco-Hillairet-Jacquier formula (\ref{E:one}) and 
our formula (\ref{E:imvol22}), one notices two main differences. First of all, formula (\ref{E:imvol22}) contains the expression $\left(U_{\frac{x_0}{K}}\right)^{-1}\left(m_T\right)$ instead of the expression ${\cal N}^{-1}\left(m_T\right)$ appearing in formula (\ref{E:one}). On the other hand, the function $\Phi$ in (\ref{E:one}) satisfies $\Phi(K)=O\left(\left(\log\frac{x_0}{K}\right)^{-\frac{1}{2}}\right)$ as $K\rightarrow 0$, while the error term in
(\ref{E:imvol22}) is $O\left(\left(\log\frac{x_0}{K}\right)^{-\frac{3}{2}}\right)$ as $K\rightarrow 0$.
\end{remark}

We will prove Theorem \ref{T:corrf} in Section \ref{S:next}, while in Section \ref{S:appl} we will derive the De Marco-Hillairet-Jacquier formula from our Theorem \ref{T:corrf}. In addition, in Section \ref{S:appl} 
we give estimates for the difference 
$$
\left(U_{\frac{x_0}{K}}\right)^{-1}\left(m_T\right)-{\cal N}^{-1}\left(m_T\right)
$$
at small strikes. Section \ref{S:CEVm} deals with
the left-wing behavior of the implied volatility in the CEV model. Finally, in the last section of the paper (Section \ref{S:num}), we compare the performance of two formulas, providing approximations to the implied volatility at small strikes in the CEV model: the De Marco-Hillairet-Jacquier formula and the formula in Corollary \ref{C:sims} in the present paper. 
\section{Proof of Theorem \ref{T:corrf}} \label{S:next}
We have already mentioned that it suffices to prove the theorem in the case where $x_0=1$.
For every small number $\varepsilon> 0$, set 
\begin{equation}
\widetilde{G}_{\varepsilon}(K)=G(K)+\frac{3{\cal N}^{-1}(m_T)^2+2+\varepsilon}{8\sqrt{\pi}(\log K)^{\frac{3}{2}}}.
\label{E:Gg}
\end{equation}
The following assertion provides two-sided estimates for the implied volatility.
\begin{theorem}\label{T:ochen}
Let $\varepsilon> 0$. Then there exists $K_{\varepsilon}> 0$ such that
\begin{align}
&\frac{\sqrt{2}}{\sqrt{T}}\sqrt{\log\frac{1}{K}+H_{2,\varepsilon}\left(\frac{1}{K}\right)} \nonumber \\
&\le I_C(K)\le\frac{\sqrt{2}}{\sqrt{T}}\sqrt{\log\frac{1}{K}+H_1\left(\frac{1}{K}\right)}
\label{E:wildi3}
\end{align}
for all $0< K< K_{\varepsilon}$. In (\ref{E:wildi3}), the functions $H_1$ and $H_{2,\varepsilon}$ are defined as follows:
\begin{align}
H_1(K)&=\left[(U_K)^{-1}(G(K))\right]^2 \nonumber \\
&\quad+(U_K)^{-1}(G(K))\sqrt{\left[(U_K)^{-1}(G(K))\right]^2+2\log K}
\label{E:fina7}
\end{align}
and 
\begin{align}
H_{2,\varepsilon}(K)&=\left[(U_K)^{-1}(\widetilde{G}_{\varepsilon}(K))\right]^2 \nonumber \\
&\quad+(U_K)^{-1}(\widetilde{G}_{\varepsilon}(K))
\sqrt{\left[(U_K)^{-1}(\widetilde{G}_{\varepsilon}(K))\right]^2+2\log K}.
\label{E:final2}
\end{align}
\end{theorem}

\it Proof of Theorem \ref{T:ochen}. \rm Our first goal is to find two functions $\widetilde{I}_1$ and $\widetilde{I}_2$  
satisfying the following conditions:
\begin{equation}
C_{BS}(K,\widetilde{I}_1(K))\le G(K)\le C_{BS}(K,\widetilde{I}_2(K))
\label{E:fina2}
\end{equation}
for sufficiently large values of $K$. The inequalities in (\ref{E:fina2}) will allow us to estimate the implied
volatility $I_G$ for large strikes, and hence to characterize the left-wing behavior of the implied volatility $I_C$.

Let $\varphi$ be a real function growing slower than $\log K$. The function $\varphi$ will be chosen later. Put
$$
\widetilde{I}(K)=\frac{\sqrt{2}}{\sqrt{T}}\sqrt{\log K+\varphi(K)}.
$$
Then we have
$$
d_1(K,\widetilde{I}(K))=\frac{\varphi(K)}{\sqrt{2}\sqrt{\log K+\varphi(K)}}
$$
and
$$
d_2(K,\widetilde{I}(K))=\frac{-2\log K-\varphi(K)}
{\sqrt{2}\sqrt{\log K+\varphi(K)}}.
$$
Therefore,
\begin{align}
&C_{BS}(K,\widetilde{I}(K))={\cal N}\left(d_1(K,\widetilde{I}(K))\right)
-K{\cal N}\left(d_2(K,\widetilde{I}(K))\right)
\nonumber \\
&={\cal N}\left(\frac{\varphi(K)}{\sqrt{2}\sqrt{\log K+\varphi(K)}}\right)
-K{\cal N}\left(\frac{-2\log K-\varphi(K)}{\sqrt{2}\sqrt{\log K+\varphi(K)}}\right) \nonumber \\
&=A(K)-B(K).
\label{E:ro1}
\end{align}

Our next goal is to estimate the last term in (\ref{E:ro1}). We will use the following known inequalities: 
\begin{equation}
\frac{1}{\sqrt{2\pi}}\left[\frac{1}{x}-\frac{1}{x(x^2+1)}\right]e^{-\frac{x^2}{2}}
\le\frac{1}{\sqrt{2\pi}}\int_x^{\infty}e^{-\frac{y^2}{2}}dy
\le\frac{1}{\sqrt{2\pi}}\frac{1}{x}e^{-\frac{x^2}{2}}.
\label{E:sestim1}
\end{equation}
The estimates in (\ref{E:sestim1}) follow from stronger inequalities formulated in \cite{AS}, 7.1.13. Taking into account (\ref{E:sestim1}), we see that
\begin{align}
&B(K)=K{\cal N}\left(\frac{-2\log K-\varphi(K)}{\sqrt{2}\sqrt{\log K+\varphi(K)}}\right) \nonumber \\
&\le\frac{1}{\sqrt{\pi}}\frac{\sqrt{\log K+\varphi(K)}}{2\log K+\varphi(K)}
\exp\left\{-\frac{\varphi(K)^2}{4(\log K+\varphi(K))}\right\}
\label{E:sestim2}
\end{align}
and
\begin{align}
&B(K)\ge\frac{1}{\sqrt{\pi}}\frac{\sqrt{\log K+\varphi(K)}}{2\log K+\varphi(K)}
\exp\left\{-\frac{\varphi(K)^2}{4(\log K+\varphi(K))}\right\} \nonumber \\
&\quad-\frac{2}{\sqrt{\pi}}\frac{(\log K+\varphi(K))^{\frac{3}{2}}}
{(2\log K+\varphi(K))[(2\log K+\varphi(K))^2+2(\log K+\varphi(K))]} \nonumber \\
&\quad\exp\left\{-\frac{\varphi(K)^2}{4(\log K+\varphi(K))}\right\}.
\label{E:sestim3}
\end{align}

Let us next suppose that the function $\varphi$ has the following form:
\begin{equation}
\varphi(K)=\Phi(K)\sqrt{\log K},
\label{E:phi}
\end{equation}
where $\Phi$ is such that 
\begin{equation}
\lim_{K\rightarrow\infty}\Phi(K)=\gamma.
\label{E:constant}
\end{equation}
In (\ref{E:constant}), $\gamma$ is a real number. Then we have
\begin{align}
&0\le\frac{1}{2\sqrt{\log K}}-\frac{\sqrt{\log K+\varphi(K)}}{2\log K+\varphi(K)} \nonumber \\
&\le\frac{1}{2\sqrt{\log K}}\left[1+\frac{\Phi(K)}{2\sqrt{\log K}}-\sqrt{1+\frac{\Phi(K)}{\sqrt{\log K}}}
\right] \nonumber \\
&\le\frac{3\Phi(K)^2}{16(\log K)^{\frac{3}{2}}},
\label{E:rur}
\end{align}
for all $K> K_0$. It follows from (\ref{E:sestim2}) and the first inequality in (\ref{E:rur}) that
\begin{equation}
B(K)\le\frac{1}{2\sqrt{\pi}\sqrt{\log K}}\exp\left\{-\frac{\Phi(K)^2\log K}{4\left(\log K+\Phi(K)\sqrt{\log K}
\right)}\right\},
\label{E:cons}
\end{equation}
for all $K> K_0$. Note that
$$
\exp\left\{-\frac{\Phi(K)^2\log K}{4\left(\log K+\Phi(K)\sqrt{\log K}
\right)}\right\}\rightarrow\exp\left\{-\frac{\gamma^2}{4}\right\}
$$
as $K\rightarrow\infty$. 

To get a lower estimate for $B(K)$, we observe that
\begin{align}
&\frac{2}{\sqrt{\pi}}\frac{(\log K+\varphi(K))^{\frac{3}{2}}}
{(2\log K+\varphi(K))[(2\log K+\varphi(K))^2+2(\log K+\varphi(K))]} \nonumber \\
&\quad\exp\left\{-\frac{\varphi(K)^2}{4(\log K+\varphi(K))}\right\} \nonumber \\
&\le\frac{1}{4\sqrt{\pi}}\frac{(\log K+\varphi(K))^{\frac{3}{2}}}{(\log K)^3}
=\frac{1}{4\sqrt{\pi}}\frac{\left(1+\frac{\Phi(K)}{\sqrt{\log K}}\right)^{\frac{3}{2}}}{(\log K)^{\frac{3}{2}}}
\nonumber \\
&\le\frac{1}{4\sqrt{\pi}}\left[\frac{1}{(\log K)^{\frac{3}{2}}}+\frac{3\Phi(K)}{2(\log K)^2}+\frac{3\Phi(K)^2}
{8(\log K)^{\frac{5}{2}}}\right],\quad K> K_1.
\label{E:coc}
\end{align}
In the proof of (\ref{E:coc}), we used the inequality 
$$
(1+h)^{\frac{3}{2}}\le 1+\frac{3}{2}h+\frac{3}{8}h^2,\quad 0< h< h_0.
$$
It follows from (\ref{E:sestim3}), (\ref{E:phi}), (\ref{E:rur}), and (\ref{E:coc}) that
\begin{align*}
&B(K)\ge\frac{1}{2\sqrt{\pi}\sqrt{\log K}}\exp\left\{-\frac{\Phi(K)^2\log K}{4\left(\log K+\Phi(K)\sqrt{\log K}
\right)}\right\} \\
&\quad-\left[
\frac{3\Phi(K)^2+4}{16\sqrt{\pi}(\log K)^{\frac{3}{2}}}+\frac{3\Phi(K)}{8\sqrt{\pi}(\log K)^2}
+\frac{3\Phi(K)^2}{32\sqrt{\pi}(\log K)^{\frac{5}{2}}}\right],
\end{align*}
for all $K> K_2$. Therefore, condition (\ref{E:constant}) implies that for every $\varepsilon> 0$ there
exists a sufficiently large number
$K_{\varepsilon}> 0$ such that for every $K> K_{\varepsilon}$,
\begin{align*}
B(K)&\ge\frac{1}{2\sqrt{\pi}\sqrt{\log K}}\exp\left\{-\frac{\Phi(K)^2\log K}{4\left(\log K+\Phi(K)\sqrt{\log K}
\right)}\right\} \\
&\quad-\frac{3\gamma^2+4+\varepsilon}{16\sqrt{\pi}(\log K)^{\frac{3}{2}}}.
\end{align*}

Next, taking into account (\ref{E:ro1}), (\ref{E:cons}), and the previous inequality, we see that the following statement holds.
\begin{lemma}\label{L:estime}
For every $\varepsilon> 0$ and all $K> K_{\varepsilon}$,
$$
Z(K)\le C_{BS}(K,\widetilde{I}(K))\le Z(K)+\frac{3\gamma^2+4+\varepsilon}{16\sqrt{\pi}(\log K)^{\frac{3}{2}}},
$$
where
\begin{align}
Z(K)&={\cal N}\left(\frac{\Phi(K)\sqrt{\log K}}{\sqrt{2}\sqrt{\log K+\Phi(K)\sqrt{\log K}}}\right) \nonumber \\ 
&\quad-\frac{1}{2\sqrt{\pi}\sqrt{\log K}}\exp\left\{-\frac{\Phi(K)^2\log K}{4\left(\log K+\Phi(K)\sqrt{\log K}
\right)}\right\}.
\label{E:fina}
\end{align}
\end{lemma}

Lemma \ref{L:estime} provides estimates for the function $C_{BS}(K,\widetilde{I}(K))$, which differ by a quantity of the higher order of smallness than $\frac{1}{\sqrt{\log K}}$.

Let us next choose the function $\Phi_1$ so that the function $Z_1$, given by the formula in (\ref{E:fina})
with $\Phi_1$ instead of $\Phi$, satisfies $Z_1(K)=G(K)$. It follows from (\ref{E:fina}) 
that
\begin{equation}
\frac{\Phi_1(K)\sqrt{\log K}}{\sqrt{2}\sqrt{\log K+\Phi_1(K)\sqrt{\log K}}}=(U_K)^{-1}(G(K)).
\label{E:fina4}
\end{equation}
Now, using (\ref{E:fina4}) we see that in order to find the value of $\Phi_1(K)$ we should solve the following
quadratic equation:
\begin{align}
&\sqrt{\log K}\Phi_1(K)^2-2\left[(U_K)^{-1}(G(K))\right]^2\Phi_1(K) \nonumber \\
&\quad-2\sqrt{\log K}\left[(U_k)^{-1}(G(K))\right]^2=0.
\label{E:fina5}
\end{align}
Solving (\ref{E:fina5}) and taking into account (\ref{E:fina4}), we obtain
\begin{equation}
\Phi_1(K)=\frac{H_1(K)}{\sqrt{\log K}},
\label{E:fina6}
\end{equation}
where $H_1$ is defined by (\ref{E:fina7}).

We will next check that the function $\Phi_1$ defined by (\ref{E:fina6}) is admissible, that is, 
condition (\ref{E:constant}) holds for $\Phi_1$. Recall that $Z_1(K)=G(K)$, and hence (\ref{E:r2})
gives 
$$
\lim_{K\rightarrow\infty}Z_1(K)=m_T.
$$ 
It follows from the definition of the function $Z_1$
(see (\ref{E:fina})) that
$$
\lim_{K\rightarrow\infty}{\cal N}
\left(\frac{\Phi_1(K)\sqrt{\log K}}{\sqrt{2}\sqrt{\log K+\Phi_1(K)\sqrt{\log K}}}\right)=m_T.
$$
Thus
\begin{equation}
\lim_{K\rightarrow\infty}
\frac{\Phi_1(K)\sqrt{\log K}}{\sqrt{2}\sqrt{\log K+\Phi_1(K)\sqrt{\log K}}}={\cal N}^{-1}(m_T).
\label{E:ogran1}
\end{equation}

It is not hard to see that $\Phi_1$ is a bounded function. Indeed, (\ref{E:ogran1}) implies that
$$
\frac{|\Phi_1(K)|\sqrt{\log K}}{\sqrt{2}\sqrt{\log K+\Phi_1(K)\sqrt{\log K}}}< M,
$$
where $M$ is a positive constant. Therefore,
$$
\frac{1}{\sqrt{2}M}<\sqrt{\frac{1}{\Phi_1(K)^2}+\frac{1}{\Phi_1(K)\sqrt{\log K}}},
$$
and thus the function $\Phi_1$ is bounded for large values of $K$. Now, using (\ref{E:ogran1}), we see
that
$$
\lim_{K\rightarrow\infty}\Phi_1(K)=\sqrt{2}{\cal N}^{-1}(m_T),
$$
and hence $\gamma_1=\sqrt{2}{\cal N}^{-1}(m_T)$, where $\gamma_1$ is the constant appearing in (\ref{E:constant}) for the function $\Phi_1$.

It follows from Lemma \ref{L:estime} and the equality $Z_1(K)=G(K)$ that
$$
C_{BS}(K,I_G(K))\le C_{BS}(K,\widetilde{I}_1(K)),
$$
where
\begin{align}
&\widetilde{I}_1(K)=\frac{\sqrt{2}}{\sqrt{T}}\sqrt{\log K+H_1(K)},
\label{E:wild1}
\end{align}
and $H_1(K)$ is given by (\ref{E:fina7}). Therefore,
\begin{equation}
I_G(K)\le \widetilde{I}_1(K),\quad K> K_0.
\label{E:wild2}
\end{equation}

To get a lower estimate for $I_G$, we will reason similarly. The only difference here is that we replace the equation
$Z_1(K)=G(K)$ by the equation $Z_{2,\varepsilon}(K)=\widetilde{G}_{\varepsilon}(K)$ ,
where $\widetilde{G}_{\varepsilon}$ is defined by (\ref{E:Gg}),
$\varepsilon> 0$ is fixed, and $Z_{2,\varepsilon}$ is a function such as in (\ref{E:fina}), but with an unknown function
$\Phi_{2,\varepsilon}$ instead of the function $\Phi$. Next, we can prove that
$$
\Phi_{2,\varepsilon}(K)=\frac{H_{2,\varepsilon}(K)}{\sqrt{\log K}},
$$
where the function $H_{2,\varepsilon}$ is given by (\ref{E:final2}).
Moreover, the function $\Phi_{2,\varepsilon}$ is bounded for large values of $K$, and
$$
\lim_{K\rightarrow\infty}\Phi_{2,\varepsilon}(K)=\sqrt{2}{\cal N}^{-1}(m_T).
$$
Therefore $\gamma_2=\sqrt{2}{\cal N}^{-1}(m_T)$, where $\gamma_2$ is the constant appearing in (\ref{E:constant}) for the function $\Phi_{2,\varepsilon}$.

Let us set
\begin{align}
&\widetilde{I}_{2,\varepsilon}(K)=\frac{\sqrt{2}}{\sqrt{T}}\sqrt{\log K+H_{2,\varepsilon}(K)},
\label{E:wildi1}
\end{align}
with $H_{2,\varepsilon}(K)$ given by (\ref{E:final2}). It follows that
$$
C_{BS}(K,\widetilde{I}_{2,\varepsilon}(K))\le C_{BS}(K,I_G(K)),
$$
and hence
\begin{equation}
\widetilde{I}_{2,\varepsilon}(K)\le I_G(K),\quad K> K_{1,\varepsilon}.
\label{E:wildi2}
\end{equation}
Now, it is clear that formula (\ref{E:wildi3}) in Theorem \ref{T:ochen} follows from (\ref{E:r1}), (\ref{E:wild1}), (\ref{E:wild2}), (\ref{E:wildi1}), and (\ref{E:wildi2}).

This completes the proof of Theorem \ref{T:ochen}.

We will next estimate the difference between the lower and the upper estimates for the implied volatility
$I_C$ in formula (\ref{E:wildi3}). First, note that since the function $\Phi_1$ is eventually bounded, formula
(\ref{E:fina4}) shows that the function $K\mapsto(U_K)^{-1}(G(K))$ is also eventually bounded. Similarly,
the function $K\mapsto(U_K)^{-1}(\widetilde{G}_{\varepsilon}(K))$ is eventually bounded. It follows from (\ref{E:fina7})
and (\ref{E:final2}) that the functions $K\mapsto|H_1(K)|$ and $K\mapsto|H_{2,\varepsilon}(K)|$, are equivalent to the function $\sqrt{\log K}$ near infinity. Therefore,
\begin{align}
&\sqrt{\log K+H_1(K)}-\sqrt{\log K+H_{2,\varepsilon}(K)} \nonumber \\
&=\frac{H_1(K)-H_{2,\varepsilon}(K)}{\sqrt{\log K+H_1(K)}
+\sqrt{\log K+H_{2,\varepsilon}(K)}} \nonumber \\
&=O\left(\frac{H_1(K)-H_{2,\varepsilon}(K)}{\sqrt{\log K}}\right),
\label{E:smale}
\end{align}
as $K\rightarrow\infty$.

To estimate the difference $H_1(K)-H_{2,\varepsilon}(K)$, we observe that the function $(U_K)^{-1}$ is Lipschitz on every proper subinterval of the interval $(U_K(-\sqrt{2}\sqrt{\log K}),1)$ with the Lipschitz constant independent of $K$ on every such interval. Now, it is not hard to see, using (\ref{E:Gg}), (\ref{E:fina7}), and (\ref{E:final2}) that
$$
H_1(K)-H_{2,\varepsilon}(K)=O\left(\sqrt{\log K}[\widetilde{G}_{\varepsilon}(K)-G(K)]\right)=O\left(\frac{1}{\log K}\right)
$$
as $K\rightarrow\infty$. Next, by taking into account (\ref{E:smale}), we obtain
\begin{equation}
\sqrt{\log K+H_1(K)}-\sqrt{\log K+H_{2,\varepsilon}(K)}=O\left((\log K)^{-\frac{3}{2}}\right)
\label{E:imvol}
\end{equation}
as $K\rightarrow\infty$. 
\begin{theorem}\label{T:imvol}
The following formula holds for the implied volatility $I_C(K)$ as $K\rightarrow 0$:
\begin{equation}
I_C(K)=\frac{\sqrt{2}}{\sqrt{T}}\sqrt{\log\frac{1}{K}+H_1\left(\frac{1}{K}\right)}
+O\left(\left(\log\frac{1}{K}\right)^{-\frac{3}{2}}\right).
\label{E:imvol1}
\end{equation}
In (\ref{E:imvol1}), the function $H_1$ is defined by (\ref{E:fina7}).
\end{theorem}

Theorem \ref{T:imvol} follows from Theorem \ref{T:ochen} and (\ref{E:imvol}).
\begin{remark}\label{R:1} \rm Note that the $O$-estimate in formula (\ref{E:imvol1}) depends on $\varepsilon$.
More precisely, formula (\ref{E:imvol1}) should be understood as follows. For every $\varepsilon> 0$ there
exist $c_{\varepsilon}> 0$ and $K_{\varepsilon}> 0$ such that
$$
0\le I_C(K)-\frac{\sqrt{2}}{\sqrt{T}}\sqrt{\log\frac{1}{K}+H_1\left(\frac{1}{K}\right)}
\le c_{\varepsilon}\left(\log\frac{1}{K}\right)^{-\frac{3}{2}}
$$
for all $0<K< K_{\varepsilon}$.
\end{remark}

\it Proof of Theorem \ref{T:corrf} (continuation). \rm Expanding the function $x\mapsto\sqrt{1+x}$ near zero, we obtain
\begin{equation}
\sqrt{1+x}=1+\frac{1}{2}x-\frac{1}{8}x^2+\frac{1}{16}x^3+O(x^4)
\label{E:expan}
\end{equation}
as $x\rightarrow 0$. Now, using (\ref{E:expan}) and the fact that the function $K\mapsto |H_1(K)|$ grows like 
$\sqrt{\log K}$, we see that
\begin{align}
\sqrt{\log K+H_1(K)}&=\sqrt{\log K}\sqrt{1+\frac{H_1(K)}{\log K}} \nonumber \\
&=\sqrt{\log K}+\frac{H_1(K)}{2(\log K)^{\frac{1}{2}}}-\frac{H_1(K)^2}{8(\log K)^{\frac{3}{2}}} \nonumber \\
&\quad+\frac{H_1(K)^3}{16(\log K)^{\frac{5}{2}}}+O\left((\log K)^{-\frac{3}{2}}\right)
\label{E:ex1}
\end{align}
as $K\rightarrow\infty$. Moreover, (\ref{E:fina7}) and (\ref{E:expan}) imply that
\begin{align}
&H_1(K)=\left[(U_K)^{-1}(G(K))\right]^2 \nonumber \\
&\quad+\sqrt{2}{\sqrt{\log K}}(U_K)^{-1}(G(K))\sqrt{1+\frac{\left[(U_K)^{-1}(G(K))\right]^2}
{2\log K}} \nonumber \\
&=\sqrt{2}(U_K)^{-1}(G(K))\sqrt{\log K}+\left[(U_K)^{-1}(G(K))\right]^2 \nonumber \\
&\quad+\frac{\sqrt{2}\left[(U_K)^{-1}(G(K))\right]^3}{4\sqrt{\log K}}
+O\left((\log K)^{-\frac{3}{2}}\right)
\label{E:imvol3}
\end{align}
as $K\rightarrow\infty$. Next, combining (\ref{E:ex1}) and (\ref{E:imvol3}), we see that
\begin{align}
&\sqrt{\log K+H_1(K)}=(\log K)^{\frac{1}{2}}+\frac{\sqrt{2}}{2}(U_K)^{-1}(G(K))
+\frac{\left[(U_K)^{-1}(G(K))\right]^2}{2(\log K)^{\frac{1}{2}}} \nonumber \\
&+\frac{\sqrt{2}\left[(U_K)^{-1}(G(K))\right]^3}{8\log K}-\frac{\left[(U_K)^{-1}(G(K))\right]^2}
{4(\log K)^{\frac{1}{2}}}-\frac{\sqrt{2}\left[(U_K)^{-1}(G(K))\right]^3}{4\log K} \nonumber \\
&+\frac{\sqrt{2}\left[(U_K)^{-1}(G(K))\right]^3}{8\log K}+O\left((\log K)^{-\frac{3}{2}}\right)
\nonumber \\
&=(\log K)^{\frac{1}{2}}+\frac{\sqrt{2}}{2}(U_K)^{-1}(G(K))+\frac{\left[(U_K)^{-1}(G(K))\right]^2}{4}(\log K)^{-\frac{1}{2}} \nonumber \\
&+O\left((\log K)^{-\frac{3}{2}}\right)
\label{E:imvol4}
\end{align}
as $K\rightarrow\infty$. Now, it is clear that (\ref{E:imvol2}) follows from (\ref{E:imvol1}) and (\ref{E:imvol4}).

This completes the proof of Theorem \ref{T:corrf}.
\section{Corollaries}\label{S:appl}
In the present section, we explain how to derive the asymptotic formula for the left wing of the implied volatility due to De Marco, Hillairet, and Jacquier from our formula (\ref{E:imvol2}). Note that formula (\ref{E:imvol2}) 
is very sensitive to even small changes. Such changes often produce errors of order $O\left((\log K)^{-\frac{1}{2}}\right)$
as $K\rightarrow 0$.

The next statement is essentially the result obtained in Theorem 3.7 in \cite{DMHJ}.
\begin{corollary}\label{C:fina}
Let $x_0> 0$. Then 
\begin{align}
&I_C(K)=\frac{\sqrt{2}}{\sqrt{T}}\left(\log\frac{x_0}{K}\right)^{\frac{1}{2}}+\frac{{\cal N}^{-1}(m_T)}{\sqrt{T}} 
+\frac{\sqrt{2}{\cal N}^{-1}(m_T)^2}{4\sqrt{T}}\left(\log\frac{x_0}{K}\right)^{-\frac{1}{2}}
\nonumber \\ 
&\quad+\Phi\left(\frac{x_0}{K}\right).
\label{E:dmhjf}
\end{align}
In (\ref{E:dmhjf}), the function $\Phi$ satisfies the following condition:
$$
\limsup_{u\rightarrow\infty}\frac{\Phi(u)}{\Psi(u)}\le 1,
$$
where
\begin{align}
&\Psi(u)=\frac{\sqrt{2}}{2\sqrt{T}}(\log u)^{-\frac{1}{2}}+\frac{\sqrt{2\pi}}{\sqrt{T}}\exp\left\{\frac{{\cal N}^{-1}(m_T)^2}{2}\right\}\psi(u).
\label{E:limsup}
\end{align}
\end{corollary} 

\it Proof. \rm
Our first goal is to replace the expression $(U_K)^{-1}(G(K))$ in formula (\ref{E:imvol2}) by the expression ${\cal N}^{-1}(m_T)$, and estimate the error. For the sake of shortness, we put 
\begin{equation}
\tau(K)=(U_K)^{-1}(G(K))-{\cal N}^{-1}(m_T).
\label{E:ppf}
\end{equation}
\begin{lemma}\label{L:ttu}
Let $x_0> 0$. Then the following asymptotic formula is valid as $K\rightarrow 0$:
\begin{align*}
&I_C(K)=\frac{\sqrt{2}}{\sqrt{T}}\left(\log\frac{x_0}{K}\right)^{\frac{1}{2}}+\frac{{\cal N}^{-1}(m_T)}{\sqrt{T}} 
+\frac{\sqrt{2}{\cal N}^{-1}(m_T)^2}{4\sqrt{T}}\left(\log\frac{x_0}{K}\right)^{-\frac{1}{2}} 
\\
&\quad
+\eta\left(\frac{x_0}{K}\right)+O\left(\left(\log\frac{x_0}{K}\right)^{-\frac{3}{2}}\right),
\end{align*}
where 
$$
\eta(u)=\frac{\tau(u)}{\sqrt{T}}+\frac{\sqrt{2}\left[2{\cal N}^{-1}(m_T)\tau(u)
+\tau(u)^2\right]}{4\sqrt{T}}(\log u)^{-\frac{1}{2}},
$$
and $\tau$ is defined by (\ref{E:ppf}).
\end{lemma}

Lemma \ref{L:ttu} follows from Theorem \ref{T:corrf} and (\ref{E:ppf}). 

The next lemma provides an estimate for the function $\tau$.
\begin{lemma}\label{L:smallch}
The following formula holds:
$$
\limsup_{K\rightarrow\infty}\frac{\tau(K)}{\Psi(K)}\le 1,
$$
where the function $\Psi$ is given by (\ref{E:limsup}).
\end{lemma}

\it Proof. \rm Let us first assume that ${\cal N}^{-1}(m_T)\ge 0$. This assumption is equivalent to the following:
$\frac{1}{2}\le m_T< 1$. Then, using (\ref{E:fina3}), we see that for 
${\cal N}^{-1}(m_T)\le x<\infty$,
we have
$$
{\cal N}(x)-\frac{1}{2\sqrt{\pi\log K}}\exp\left\{-\frac{1}{2}{\cal N}^{-1}(m_T)^2\right\}\le U_K(x)\le{\cal N}(x).
$$
Therefore, for $y\in[m_T,1)$,
\begin{align}
&{\cal N}^{-1}(y)\le(U_K)^{-1}(y) \nonumber \\
&\le{\cal N}^{-1}\left(y+\frac{1}{2\sqrt{\pi\log K}}
\exp\left\{-\frac{1}{2}{\cal N}^{-1}(m_T)^2\right\}\right)
\label{E:itfoo}
\end{align}
and
\begin{align*}
{\cal N}^{-1}(G(K))&\le(U_K)^{-1}(G(K)) \\
&\le{\cal N}^{-1}\left(G(K)+\frac{1}{2\sqrt{\pi\log K}}\exp\left\{-\frac{1}{2}{\cal N}^{-1}(m_T)^2\right\}\right).
\end{align*}
Since
$$
{\cal N}^{-1}(y)^{\prime}=\frac{1}{{\cal N}^{\prime}\left({\cal N}^{-1}(y)\right)}=\sqrt{2\pi}
\exp\left\{\frac{{\cal N }^{-1}(y)^2}{2}\right\},
$$
the mean value theorem implies that
\begin{equation}
(U_K)^{-1}(G(K))={\cal N}^{-1}(G(K))+T(K),
\label{E:l1}
\end{equation}
where
\begin{align*}
&0\le T(K)\le\frac{\sqrt{2}}{2\sqrt{\log K}}\exp\left\{-\frac{1}{2}{\cal N }^{-1}(m _T)^2\right\} \\
&\exp\left\{\frac{1}{2}{\cal N}^{-1}\left(G(K)+\frac{1}{2\sqrt{\pi}\sqrt{\log K}}
\exp\left\{-\frac{1}{2}{\cal N}^{-1}(m_T)^2\right\}\right)^2\right\}.
\end{align*}
Next, using 
(\ref{E:r2}) and (\ref{E:m}),  we see that
for every $\varepsilon > 0$ there exists $K_{\varepsilon}> 0$ such that
\begin{equation}
T(K)\le\frac{\sqrt{2}}{2\sqrt{\log K}}(1+\varepsilon).
\label{E:ll}
\end{equation}
Moreover, (\ref{E:r2}), (\ref{E:m}), and the mean value theorem imply that
\begin{equation}
{\cal N}^{-1}(G(K))={\cal N}^{-1}(m_T)+\rho(K),
\label{E:pk}
\end{equation}
where the function $\rho$ is positive and satisfies the following condition:
\begin{equation}
\rho(K)\le\sqrt{2\pi}\psi(K)\exp\left\{\frac{1}{2}{\cal N}^{-1}\left(m_T\right)^2+\varepsilon\right\}.
\label{E:pkk}
\end{equation}

Next, taking into account formulas (\ref{E:l1}) - (\ref{E:pkk}),
we see that Lemma \ref{L:smallch} holds under the condition $\frac{1}{2}\le m_T< 1$. 

It remains to prove Lemma \ref{L:smallch} in the case where $0< m_T<\frac{1}{2}$. The previous condition means that 
${\cal N}^{-1}(m_T)< 0$. Fix $\delta> 0$ such that ${\cal N}^{-1}(m_T+\delta)< 0$. In addition, fix $K$ so large that
the following inequalities hold:
\begin{equation}
-\sqrt{2}\sqrt{\log K}<{\cal N}^{-1}(G(K)),
\label{E:i1}
\end{equation}
\begin{equation}
{\cal N}^{-1}(G(K)+\delta)< 0,
\label{E:i2}
\end{equation}
and
\begin{equation}
\frac{1}{2\sqrt{\pi\log K}}
\exp\left\{-\frac{1}{2}{\cal N}^{-1}(G(K)+\delta)^2\right\}<\delta.
\label{E:inver1}
\end{equation}
Next, taking into account (\ref{E:i1}), we assume that
$$
-\sqrt{2}\sqrt{\log K}\le x\le{\cal N}^{-1}(G(K)+\delta).
$$
Then, using (\ref{E:i2}), we obtain
$$
{\cal N}(x)-\frac{1}{2\sqrt{\pi\log K}}\exp\left\{-\frac{1}{2}{\cal N}^{-1}(G(K)+\delta)^2
\right\}\le U_K(x)\le{\cal N}(x).
$$
In addition,
\begin{align*}
&{\cal N}^{-1}(y)\le(U_K)^{-1}(y) \\
&\le{\cal N}^{-1}\left(y+\frac{1}{2\sqrt{\pi\log K}}
\exp\left\{-\frac{1}{2}{\cal N}^{-1}(G(K)+\delta)^2\right\}\right),
\end{align*}
provided that
$$
{\cal N}(-\sqrt{2}\sqrt{\log K})\le y\le G(K)+\delta-\frac{1}{2\sqrt{\pi\log K}}
\exp\left\{-\frac{1}{2}{\cal N}^{-1}(G(K)+\delta)^2\right\}.
$$

It follows from (\ref{E:i1}) and (\ref{E:inver1}) that the number $y=G(K)$ satisfies the previous condition. Therefore,
\begin{align}
&{\cal N}^{-1}(G(K))\le(U_K)^{-1}(G(K)) \nonumber \\
&\le{\cal N}^{-1}\left(G(K)+\frac{1}{2\sqrt{\pi\log K}}
\exp\left\{-\frac{1}{2}{\cal N}^{-1}(G(K)+\delta)^2\right\}\right).
\label{E:itfo}
\end{align}
Moreover, (\ref{E:itfo}) and the mean value theorem imply that for $K> K_{\delta}$,
$$
0\le T(K)\le\frac{\sqrt{2}}{2\sqrt{\log K}}\exp\left\{V(K,\delta)\right\},
$$
where $T(K)$ is determined from (\ref{E:l1}), and
\begin{align*}
V(K,\delta)&=\frac{1}{2}{\cal N}^{-1}\left(G(K)+\frac{1}{2\sqrt{\pi}\sqrt{\log K}}
\exp\left\{-\frac{1}{2}{\cal N}(G(K)+\delta)^2\right\}\right)^2 \\
&\quad-\frac{1}{2}{\cal N }^{-1}(G(K)+\delta)^2.
\end{align*}

It is easy to see that
$$
\lim_{\delta\rightarrow 0}\lim_{K\rightarrow\infty}V(K,\delta)=0.
$$ 
Therefore, for every $\varepsilon > 0$ there exists $K_{\varepsilon}> 0$ such that the inequality in 
(\ref{E:ll}) holds for all $K> K_{\varepsilon}$. Now, the proof of Lemma \ref{L:smallch} in the
case where $0< m_T<\frac{1}{2}$ can be completed exactly as in the case when $\frac{1}{2}\le m_T< 1$.

Finally, it is not hard to see that Corollary \ref{C:fina} follows from Lemmas \ref{L:ttu} and \ref{L:smallch}.

Let us assume $x_0=1$ and $0< K< 1$. We will next compare the numbers $\left(U_{\frac{1}{K}}\right)^{-1}(m_t)$ and 
${\cal N}^{-1}(m_T)$ appearing in formula (\ref{E:imvol22}) in Corollary \ref{C:sims} and in the De Marco-Hillairet-Jacquier formula (\ref{E:dmhjf}), respectively. Recall that if $\frac{1}{2}\le m_T< 1$, then
the number $\left(U_{\frac{1}{K}}\right)^{-1}(m_T)$ is defined for all $0< K< 1$. On the other hand, if $0< m_T<\frac{1}{2}$,
then $\left(U_{\frac{1}{K}}\right)^{-1}(m_T)$ is defined under the additional restriction
$$
U_{\frac{1}{K}}\left(-\sqrt{2}\sqrt{\log\frac{1}{K}}\right)< m_T.
$$

It is clear that if $\frac{1}{2}< m_T< 1$, then $\left(U_{\frac{1}{K}}\right)^{-1}(m_T)$ and ${\cal N}^{-1}(m_T)$  
are positive numbers for all $0< K< 1$. If $m_T=\frac{1}{2}$, then we have 
$\left(U_{\frac{1}{K}}\right)^{-1}(m_T)> 0$ and ${\cal N}^{-1}(m_T)=0$. 
The remaining case where 
\begin{equation}
U_{\frac{1}{K}}\left(-\sqrt{2}\sqrt{\log\frac{1}{K}}\right)< m_T<\frac{1}{2}
\label{E:spec1}
\end{equation}
is interesting. In this case, the number ${\cal N}^{-1}(m_T)$ is negative, while the sign of the number 
$\left(U_{\frac{1}{K}}\right)^{-1}(m_T)$ can be positive or negative. We will next clarify the previous statement.
\begin{lemma}\label{L:moreint}
Suppose condition (\ref{E:spec1}) holds. Then the following are true:
\begin{enumerate}
\item Let the number $K< 1$ be such that
$$
\frac{1}{2}-\frac{1}{2\sqrt{\pi}\sqrt{\log\frac{1}{K}}}< m_T<\frac{1}{2}.
$$
Then $\left(U_{\frac{1}{K}}\right)^{-1}(m_T)> 0$.
\item Let the number $K< 1$ be such that
$$
\frac{1}{2}-\frac{1}{2\sqrt{\pi}\sqrt{\log\frac{1}{K}}}=m_T.
$$
Then $\left(U_{\frac{1}{K}}\right)^{-1}(m_T)=0$.
\item Let the number $K< 1$ be such that
$$
U_{\frac{1}{K}}\left(-\sqrt{2}\sqrt{\log\frac{1}{K}}\right)<m_T< \frac{1}{2}-\frac{1}{2\sqrt{\pi}\sqrt{\log\frac{1}{K}}}.
$$
Then $\left(U_{\frac{1}{K}}\right)^{-1}(m_T)< 0$.
\end{enumerate}
\end{lemma}
\begin{remark}\label{R:price} \rm
Note that in the case described in part (1) of Lemma \ref{L:moreint}, the numbers $\left(U_{\frac{1}{K}}\right)^{-1}(m_T)$
and ${\cal N}^{-1}(m_T)$ have opposite signs.
\end{remark}

The proof of Lemma \ref{L:moreint} is simple, and we leave it as an exercise for the reader.

The next assertion characterizes the limiting behavior of the difference 
$\left(U_{\frac{1}{K}}\right)^{-1}(m_T)-{\cal N}^{-1}(m_T)$.
\begin{theorem}\label{T:estims}
Let $0< m_T< 1$. Then
\begin{equation}
\lim_{K\rightarrow 0}\sqrt{2}\sqrt{\log\frac{1}{K}}\left[\left(U_{\frac{1}{K}}\right)^{-1}(m_T)-{\cal N}^{-1}(m_T)\right]=1.
\label{E:forest}
\end{equation}
\end{theorem}
\begin{remark}\label{R:u} \rm For $0< m_T<\frac{1}{2}$, the expression $\left(U_{\frac{1}{K}}\right)^{-1}(m_T)$
exists if 
\begin{equation}
{\cal N}\left(-\sqrt{2}\sqrt{\log\frac{1}{K}}\right)< m_T
\label{E:suf}
\end{equation}
(see formula (\ref{E:accord})). Note that for every fixed $m_T$ with 
$0< m_T<\frac{1}{2}$, condition (\ref{E:suf}) holds for sufficiently small values of $K$. This explains how we should understand formula (\ref{E:forest}) for $0< m_T<\frac{1}{2}$. 
\end{remark}

\it Proof of Theorem \ref{T:estims}. \rm Suppose $\frac{1}{2}\le m_T< 1$, and set 
\begin{equation}
A_K=\left(U_{\frac{1}{K}}\right)^{-1}(m_T),\quad A={\cal N}^{-1}(m_T),
\label{E:notka1}
\end{equation}
and
\begin{equation}
B_K={\cal N}^{-1}\left(m_T+\frac{1}{2\sqrt{\pi\log\frac{1}{K}}}
\exp\left\{-\frac{1}{2}A^2\right\}\right).
\label{E:notka2}
\end{equation}
\begin{lemma}\label{L:aux}
For all $0< K< 1$,
\begin{align*}
\frac{1}
{\sqrt{2}\sqrt{\log\frac{1}{K}}+B_K}&\le A_K-A \\
&\le\frac{1}
{\sqrt{2}\sqrt{\log\frac{1}{K}}+A}\exp\left\{\frac{B_K^2-A^2}{2}\right\},
\end{align*}
where $A$, $A_K$, and $B_K$ are given by (\ref{E:notka1}) and (\ref{E:notka2}).
\end{lemma}

\it Proof of Lemma \ref{L:aux}. \rm Using the mean value theorem, we see that
\begin{equation}
A_K-A=\frac{m_T-U_{\frac{1}{K}}(A)}
{U_{\frac{1}{K}}^{\prime}(\theta)},
\label{E:mvt1}
\end{equation}
where $A<\theta< A_K$. It follows from (\ref{E:mvt1}) that
\begin{equation}
A_K-A=
\frac{1}
{\sqrt{2}\sqrt{\log\frac{1}{K}}+\theta}\exp\left\{\frac{\theta^2-A^2}{2}\right\}.
\label{E:thef1}
\end{equation}
Now, the estimates in Lemma \ref{L:aux} follow from (\ref{E:itfoo}) and (\ref{E:thef1}).

Let us continue the proof of Theorem \ref{T:estims}. Lemma \ref{L:aux} implies that
\begin{align}
\frac{\sqrt{2}\sqrt{\log\frac{1}{K}}+A}
{\sqrt{2}\sqrt{\log\frac{1}{K}}+B_K}  &\le(A_K-A)\left[\sqrt{2}\sqrt{\log\frac{1}{K}}+A\right] \nonumber \\
&\le\exp\left\{\frac{B_K^2-A^2}{2}\right\}.
\label{E:thef2}
\end{align}

Since $B_K\rightarrow A$ as $K\rightarrow 0$, formula (\ref{E:forest}) follows from (\ref{E:thef2}).

The remaining part of the proof of Theorem \ref{T:estims} resembles that  of the second part of Lemma
\ref{L:smallch}. Let us assume $0< m_T<\frac{1}{2}$. Then we have ${\cal N}^{-1}(m_T)< 0$. Fix $\delta> 0$ such that
\begin{equation}
{\cal N}^{-1}(m_T+\delta)< 0,
\label{E:ess0}
\end{equation} 
and suppose $K< 1$ is such that
\begin{equation}
-\sqrt{2}\sqrt{\log\frac{1}{K}}<{\cal N}^{-1}(m_T)
\label{E:ess1}
\end{equation} 
and
\begin{equation}
\frac{1}{2\sqrt{\pi}\sqrt{\log\frac{1}{K}}}\exp\left\{-\frac{1}{2}{\cal N}^{-1}(m_T+\delta)^2\right\}<\delta.
\end{equation}
Let 
\begin{equation}
-\sqrt{2}\sqrt{\log\frac{1}{K}}\le x\le{\cal N}^{-1}(m_T+\delta).
\label{E:ess2}
\end{equation}
Then we have
\begin{equation}
{\cal N}(x)-\frac{1}{2\sqrt{\pi}\sqrt{\log\frac{1}{K}}}\exp\left\{-\frac{1}{2}{\cal N}^{-1}(m_T+\delta)^2\right\}\le 
U_{\frac{1}{K}}(x)\le{\cal N}(x).
\label{E:ess3}
\end{equation}
Therefore,
\begin{align*}
&{\cal N}^{-1}(y)\le \left(U_{\frac{1}{K}}\right)^{-1}(y) \\
&\le{\cal N}^{-1}\left(y+\frac{1}{2\sqrt{\pi\log\frac{1}{K}}}
\exp\left\{-\frac{1}{2}{\cal N}^{-1}(m_T+\delta)^2\right\}\right),
\end{align*}
provided that
$$
{\cal N}\left(-\sqrt{2}\sqrt{\log\frac{1}{K}}\right)< y< m_T+\delta-\frac{1}{2\sqrt{\pi\log\frac{1}{K}}}
\exp\left\{-\frac{1}{2}{\cal N}^{-1}(m_T+\delta)^2\right\}.
$$

Since the previous estimates hold for the number $y=m_T$, we have
$A< A_K< B_{K,\delta}$. Here $A$ and $A_K$ are defined by (\ref{E:notka1}), and
$$
B_{K,\delta}={\cal N}^{-1}\left(m_T+\frac{1}{2\sqrt{\pi\log\frac{1}{K}}}
\exp\left\{-\frac{1}{2}{\cal N}^{-1}(m_T+\delta)^2\right\}\right).
$$
Next, using (\ref{E:ess0}) and (\ref{E:ess2}), we obtain 
$A< A_K< B_{K,\delta}< 0$.

It follows from (\ref{E:thef1}) and from the inequalities 
$A<\theta< B_{K,\delta}< 0$
that
$$
\frac{1}{\sqrt{2}\sqrt{\log\frac{1}{K}}+B_{K,\delta}}\exp\left\{\frac{B_{K,\delta}^2-A^2}{2}\right\}\le A_K-A
\le\frac{1}{\sqrt{2}\sqrt{\log\frac{1}{K}}+A}.
$$
Finally, it is not hard to see that formula (\ref{E:forest}) with $0< m_T<\frac{1}{2}$ can be derived from the previous estimates and from the equality 
$\lim_{K\rightarrow 0}B_{K,\delta}=A$.
\begin{remark}\label{R:eshchio} \rm
Theorem \ref{T:estims} explains why the error term in the De Marco-Hillairet-Jacquier formula is worse than that
in formula (\ref{E:imvol22}).
\end{remark}
\section{The CEV model}\label{S:CEVm}
The constant elasticity of variance model (the CEV model) is described by the following stochastic differential equation:
$$
dS_t=\sigma S_t^{\rho}dW_t,
$$
where $0<\rho< 1$, $\sigma> 0$, and $S_0=s_0$. If $\frac{1}{2}\le\rho< 1$, then the boundary at $x=0$ is naturally absorbing, while for $0<\rho<\frac{1}{2}$, we impose an absorbing boundary condition. The CEV model was introduced by
J. C. Cox in \cite{C} (see also \cite{CR}). A useful information about the CEV model, including some of the results formulated below, can be found in \cite{BL}. The CEV process is used in the financial industry to model spot prices of equitites and commodities (see, e.g., \cite{DVN,GS}, and the references therein).

Fix $T> 0$. Then we have
\begin{equation}
m_T=1-\Gamma\left(\frac{1}{2(1-\rho)},\frac{s_0^{2(1-\rho)}}{2T\sigma^2(1-\rho)^2}\right).
\label{E:cev1}
\end{equation}
Moreover, the density of the absolutely continuous part of the distribution $\widetilde{\mu}_T$ 
of $S_T$ is as follows:
\begin{align}
&\widetilde{D}_T(x) \nonumber \\
&=cx^{\frac{1}{2}-2\rho}\exp\left\{-\frac{x^{2(1-\rho)}}{2T\sigma^2(1-\rho)^2}\right\}
I_{-\nu}\left(\frac{s_0^{1-\rho}x^{1-\rho}}{T\sigma^2(1-\rho)^2}\right).
\label{E:cev2}
\end{align}
In (\ref{E:cev1}), $\Gamma$ is the normalized incomplete gamma function given by
$$
\Gamma(a,y)=\frac{1}{\Gamma(a)}\int_0^yt^{a-1}e^{-t}dt,\quad a> 0,\quad y\ge 0,
$$
while in (\ref{E:cev2}), the parameter $\nu$ is defined by $\nu=-\frac{1}{2(1-\rho)}$, the function $I_{-\nu}$ is the modified Bessel function of the first kind,
and the constant $c$ is given by
$$
c=\frac{s_0^{\frac{1}{2}}}{T\sigma^2(1-\rho)}\exp\left\{-\frac{s_0^{2(1-\rho)}}{2T\sigma^2(1-\rho)^2}\right\}.
$$

It is known that as $x\rightarrow 0$,
$$
I_{\alpha}(x)\sim\frac{1}{\Gamma(\alpha+1)}\left(\frac{x}{2}\right)^{\alpha}
$$
for all $\alpha\neq-1,-2,\cdots$. Therefore, it follows from (\ref{E:cev2}) that as $x\rightarrow 0$,
\begin{equation}
\widetilde{D}_T(x)\sim \tilde{c}x^{1-2\rho},
\label{E:cev3}
\end{equation}
where
\begin{align}
\tilde{c}&=
\frac{s_0}{T\sigma^2(1-\rho)[2T\sigma^2(1-\rho)^2]^{\frac{1}{2(1-\rho)}}\Gamma\left(\frac{3-2\rho}{2(1-\rho)}\right)} \nonumber \\
&\quad\exp\left\{-\frac{1}{2T\sigma^2(1-\rho)^2}\right\}.
\label{E:cev4}
\end{align}
Now, taking into account (\ref{E:cev3}) and (\ref{E:cev4}), we see that as $K\rightarrow 0$,
$$
\int_0^Kd\widetilde{\mu}_T(x)\sim\frac{\tilde{c}}{2(1-\rho)}K^{2(1-\rho)}.
$$
Therefore, condition (\ref{E:ost}) holds. Finally, applying Corollary \ref{C:sims}, 
we derive the following statement.
\begin{corollary}\label{C:cocor}
Formula (\ref{E:imvol22})
with $m_T$ given by (\ref{E:cev2}) holds for the implied volatility in the CEV model.
\end{corollary}
\begin{remark}\label{R:finn} \rm Propositions, similar to Corollary \ref{C:cocor}, can be established for many 
other models besides the CEV model, e.g., jump-to-default models, and models described by processes stopped at the first hitting time of zero. To apply Corollary \ref{C:sims} to a model with atoms, we only need to know the value of $m_T$ and to estimate the rate of decay of the asset price density near zero. Such information is provided in \cite{DMHJ} for some of the models mentioned above.
\end{remark}
\section {Numerics}\label{S:num}
The figures included in the present section illustrate the performance of two asymptotic formulas, providing approximations to the left wing of the implied volatility in the CEV model: the De Marco-Hillairet-Jacquier formula 
and formula (\ref{E:imvol22}) established in the present paper. The values of the CEV parameters in Figures 1 and 2 
are chosen as follows: $s_0=0.05$, $T=1.2$, $\beta=0.6$, and $\sigma=0.2$. Under the previous
assumptions, the value $m_T$ of the mass at zero is approximately equal to $0.0707$. 

\begin{figure}[htb]
     \begin{center}
\includegraphics[scale=0.35]{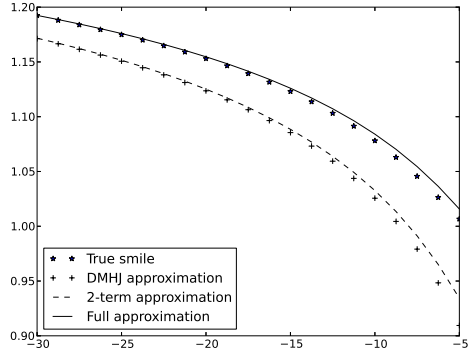}
\caption{Normalised implied volatility from formula (\ref{E:imvol22}) and the De Marco-Hillairet-Jacquier approximation ($m_T=0.0707$).}
     \end{center}
     \label{f:PlotSmiles}
\end{figure}
\begin{figure}[htb]
     \begin{center}
\includegraphics[scale=0.4]{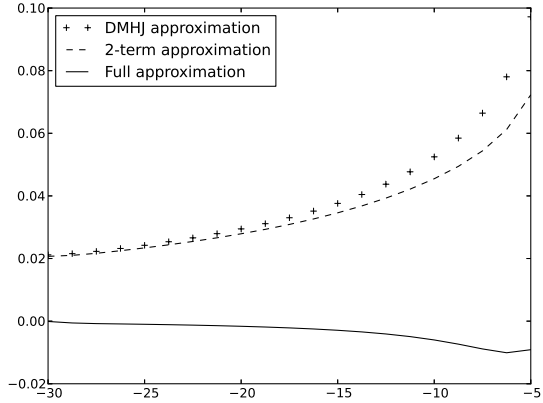}
\caption{Normalised implied volatility errors from formula (\ref{E:imvol22}) and the De Marco-Hillairet-Jacquier approximation ($m_T=0.0707$).}
     \end{center}
     \label{f:PlotSmilesErrors}
\end{figure}

In Figures 1 and 2, the independent variable is the log-moneyness $k$ given by $k=\log\frac{K}{s_0}$. The large blue stars in Figure 1 show the Monte Carlo estimate of the function 
\begin{equation}
k\mapsto I_C(k)\frac{\sqrt{T}}{|k|}.
\label{E:Mont}
\end{equation} 
To plot the graph of the function represented by blue stars, Monte Carlo simulations with $10^4$ paths were used, each drawn with 100 time steps. The solid black curve in Figure 1 depicts the full smile approximation using all the three terms in formula (\ref{E:imvol22}). Furthermore, the graph in black dashes corresponds to the smile approximation based on formula (\ref{E:imvol22}) with 2 terms, while the graph in black crosses represents the De Marco-Hillairet-Jacquier approximation. Figure 2 shows the approximation errors. 

Even superficial observations of the graphs in Figures 1 and 2 show that formula (\ref{E:imvol22}) provides a better approximation to the left wing of the implied volatility in the CEV model than the De Marco-Hillairet-Jacquier formula. Note that the graph of the Monte Carlo estimate of the function defined in (\ref{E:Mont}) and the graph of the approximation to this function based on formula (\ref{E:imvol22}) match rather well.

\end{document}